\documentclass[prl,twocolumn,superscriptaddress]{revtex4}
\usepackage{epsfig,amsmath,amssymb,graphicx,color,calc,epstopdf}

\let\f=\varphi
\newcommand{\beq}{\begin{equation}}
\newcommand{\eeq}{\end{equation}}
\newcommand{\ba}{\begin{align}}
\newcommand{\ea}{\end{align}}
\renewcommand{\phi}{\varphi}

\begin{document}

\title{Glass transition and random close packing above three dimensions}

\author{Patrick Charbonneau}
\affiliation{Departments of Chemistry and Physics, Duke University, Durham,
North Carolina 27708, USA}
\affiliation{LPTMC, CNRS-UMR 7600, Universit\'e Pierre et Marie Curie, bo\^ite 121, 4 Place Jussieu,
75005 Paris, France}

\author{Atsushi Ikeda}
\affiliation{Laboratoire Charles Coulomb,
UMR 5221 CNRS and Universit\'e Montpellier 2, Montpellier, France}

\author{Giorgio Parisi}
\affiliation{Dipartimento di Fisica,
Sapienza Universit\'a di Roma,
INFN, Sezione di Roma I, IPFC -- CNR,
P.le A. Moro 2, I-00185 Roma, Italy
}

\author{Francesco Zamponi}
\affiliation{LPT,
\'Ecole Normale Sup\'erieure, UMR 8549 CNRS, 24 Rue Lhomond, 75005 Paris, France}


\begin{abstract}
Motivated by a recently identified severe discrepancy between a static and a dynamic theory of glasses,
we numerically investigate the behavior of dense hard spheres in spatial dimensions 3 to 12.
Our results are consistent with the static replica theory, but disagree with the
dynamic mode-coupling theory, indicating that key ingredients of high-dimensional physics
are missing from the latter. We also obtain numerical
estimates of the random close packing density,
which provides new insights into the mathematical problem of packing spheres in large dimension.
\end{abstract}

\maketitle

Studying physical theories upon varying the dimensionality of space $d$
has illuminated a variety of problems over the last century.
Unifying fundamental forces through general relativity and string theories immediately come to mind, but so does the treatment of critical
phenomena through the renormalization group approach. Similar contemplations may now benefit another
grand challenge of condensed matter, that of describing glass formation~\cite{anderson:1995}. Including $d$ within the set of
control parameters nicely complements two- and three-dimensional experiments and simulations by, first, surmounting some of the
technical challenges encountered in low dimensions~\cite{charbonneau:2010}; and, second,
by bringing the problem in contact with rich mathematical fields, such as coding theory~\cite{conway:1988}.
Obtaining a complete and consistent high-dimensional theory of amorphous states would also crucially
pave the way for a more systematic understanding of finite
dimensional effects, and ultimately provide a reliable theory of the glass transition,
jamming, and related phenomena.

The Random First Order Transition (RFOT) theory has emerged as one of the main contenders in the quest for a complete description of the glass transition.
Its foundation was posed in the late eighties, when it was realized that a class of mean-field spin models
share the phenomenology of glass formers~\cite{kirkpatrick:1989}.
These models are very abstract and do not obviously present the microscopic features of particle-based systems,
but they helped turn RFOT theory into a quantitative method for studying glasses, thanks to the development of
the dynamic mode-coupling theory (MCT)~\cite{gotze:2009} and
the static replica theory (RT)~\cite{mezard:1999}.
The two quantitative theories of the glass transition are
intimately related~\cite{kirkpatrick:1987,szamel:2010}, but
make use of different approximations in order
to obtain closed-form structures~\cite{szamel:2010}. Diverging physical descriptions result.
The predictions of MCT~\cite{ikeda:2010,schmid:2010} and RT~\cite{parisi:2010} indeed strongly
disagree in the limit of large $d$, with MCT suggesting that the glass transition happens at densities
much larger than that of random close packing provided by RT. This situation is particularly worrying~\cite{bouchaud:2010}. Because RFOT theory is based on a mean-field description,
it should be exact in large $d$. If MCT is really the dynamic part of RFOT theory,
then MCT and RT should lead to identical predictions,
as proposed in~\cite{kirkpatrick:1987}. Whether MCT is really a mean-field theory, however,
has been questioned~\cite{KS91}.
In order to test the quality of the
approximations and to clarify the connection with the original mean-field models, one ought to understand the source of
the discrepancy, and to obtain a controlled limit of particle systems
in which RFOT becomes exact.

Independently from this issue, packing spheres in high dimensions
is intimately related to several important mathematical problems, notably in the
context of signal digitalization and of error correcting codes~\cite{conway:1988}.
It has been suggested that for large $d$ disordered packings may be more
efficient than lattice-based versions~\cite{torquato:2006b}.
Although lattice geometries are strongly $d$ dependent, the fluid structure and
properties have been suggested to be much less sensitive to $d$, once $d\gtrsim 3$ (see e.g. Ref.~\cite{vanmeel:2009b}).
A general understanding of disordered packings in arbitrary $d$ would thus also clarify
the density scaling of amorphous packings and their potential mathematical role.

In this letter, we compare simulations of hard spheres up to $d=12$ with
the predictions of RT and MCT. The results suggest
that while RT offers a satisfying description of dimensionality MCT fails at the task,
which calls for a modified formulation of the dynamical theory of RFOT.
The results also provide the ``random close-packed'' density in several $d$
and give a hint of the scaling of this quantity for large $d$, which should allow
comparisons with the results of other theoretical treatments~\cite{song:2008}.

\paragraph*{Numerical simulations -}
We employ a modified Lubachevsky-Stillinger
algorithm to densify a low-density gas of $N$ identical hard spheres of diameter $\sigma$, enclosed in a periodic box of volume $V$, by growing the particles
at a constant rate $\gamma = \dot{\sigma}$, reported here in standard reduced units~\cite{skoge:2006,footnote:1}.
Time evolution stops when the system reaches a high reduced pressure $p\equiv\beta P/\rho=10^3$ measured by
rescaling the mechanical pressure $P$ by the number density $\rho\equiv N/V$ and the inverse temperature $\beta$, which is
thermostated to unity. 
The packing fraction is $\f \equiv \rho V_{d} (\sigma/2)$,
where $V_{d}(R)$ is the volume of a $d$-dimensional ball of radius $R$.
Our event-driven molecular dynamics scheme complements Ref.~\cite{skoge:2006}'s
earlier implementation of cubic blocking of space with spherical nearest-neighbor lists~\cite{donev:2005b}.
Because the volume of a ball inscribed in a cube tends to zero with growing $d$, remarkable efficiency gains are obtained from considering collisions with fewer neighbors. Up to $d=10$, particles are grown at rates as low
as $\gamma=3\times10^{-5}$, while rates of $10^{-4}$ and $10^{-3}$ are attained in $d=11$ and $12$, respectively.
Systems with $N=8000$ are simulated for $d\leq9$ and larger ones for $d=10$--$12$~\cite{appendix}.
These sizes ensure that even when the system is in its densest state the box edge remains larger than $2\sigma$, which prevents a particle from ever having two direct contacts with another one. There are strong reasons to believe that although relatively small these $N$ nonetheless provide a
reliable approximation of bulk behavior. First, with increasing $d$ the box edge becomes less representative of the
overall box size. The largest diagonals are $\sqrt{d}$ larger and there are many more diagonals than edges.
Second, by analogy to spin systems, mean-field arguments indicate that for $d\to\infty$, a hypercube of side two is
sufficient to capture the full thermodynamic behavior.
Even at the critical point, finite-size corrections are proportional
to $1/N^{\delta}$, where
the exponent $\delta$ is model dependent (e.g. 1/2 at the ferromagnetic transition), and do not directly involve the edge length $L$~\cite{Pa98}. Similar results hold for dimensions greater than the upper critical dimension, where the exponents coincide with the mean field ones.
Third, the fluid structure is expected to become uniform at ever smaller distances with increasing
$d$~\cite{PS00,skoge:2006}. Nearest-neighbor ordering should thus mainly be influenced by particles in contact or nearly so,
 with the rest of the fluid acting as a continuum.
Indeed, in the fluid phase, finite volume corrections are proportional to the pair correlation $h(L)$,
and at fixed $L$, $h(L)$ goes to zero exponentially with $d$~\cite{PS00}. The validity of these rationalizations, which are consistent with the decorrelation property of high $d$ sphere
packings proposed in~\cite{torquato:2006b},
are satisfactorily tested by simulations in $d=8$~\cite{appendix}.

\begin{figure}
\includegraphics[width=\columnwidth]{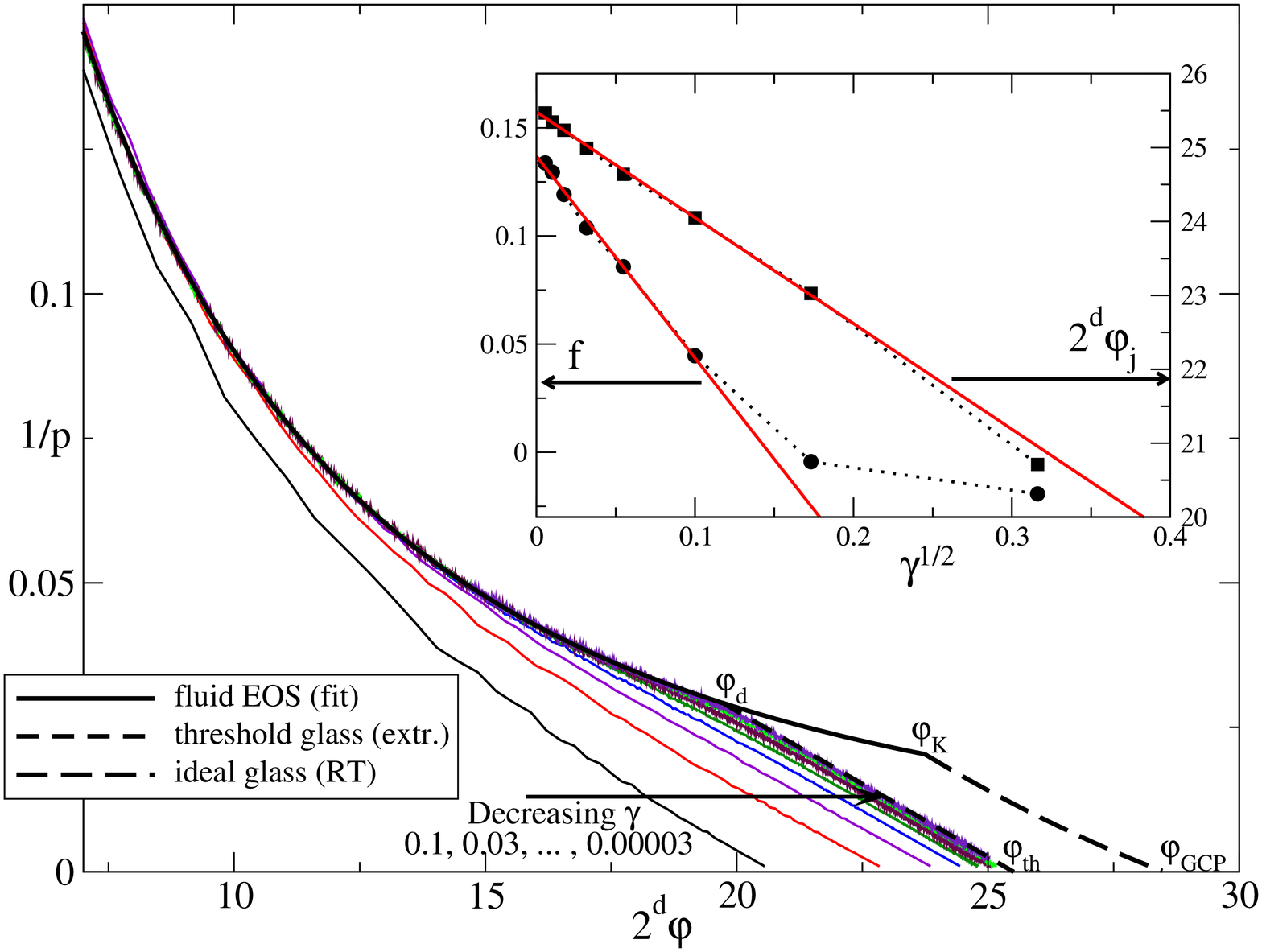}
\caption{
Different compactions of $N=8000$ particles in $d=9$.
With growing $\f$, the pressure first evolves like the fluid EOS then like a free volume EOS. Extrapolated threshold glass and theoretical ideal glass lines illustrate the subsequent analysis (see text for details).
}
\label{fig:comp}
\end{figure}

\paragraph*{Numerical results -}
The compression results for $d=9$ shown in Fig.~\ref{fig:comp}
are representative of the behavior observed for all $d>3$,
where crystallization is not observed on the time scale accessible to
present computers~\cite{skoge:2006,vanmeel:2009b};
in $d=3$ crystallization is observed at small $\gamma$ leading to a drop in the pressure
at intermediate density~\cite{skoge:2006}.
The system first follows the equilibrium fluid equation of state (EOS) at low density and falls out of
equilibrium at high density. Beyond this point, the pressure increases faster than in the equilibrium fluid and ultimately
diverges at packing fraction $\f_{\rm j}(\gamma)$.
A Carnahan-Starling form
\begin{equation}
p_{\rm fluid}(\f)=1+ 2^{d-1} \f \, \frac{1-A_{d} \, \f}{(1-\f)^d},
\label{eq:eos}
\end{equation}
captures well the pressure growth with $\f$
in the fluid regime (Fig.~\ref{fig:comp}), provided, for each $d$,
that one fits $A_{d}$ to the data from the slowest compression
rate available~\cite{footnote:2}.
Note that the coefficients $A_{d}^{\rm fit}$ are not identical
to $A_{d}^{\rm CS}$ adjusted to recover the correct third virial coefficient~\cite{SMS89} (see~\cite{appendix}), but
the values are quite close, and in any case this
contribution vanishes with increasing $d$.

In the high-density non-equilibrium regime, compaction runs with different $\gamma$ follow separate branches
along which the pressure evolution is dominated by the expulsion of free volume~\cite{donev:2005,kamien:2007}.
Upon approaching jamming, the pressure is well approximated by
\begin{equation}
p_{\rm fv}(\gamma,\f)=\frac{d \, \f_{\mathrm{j}}(\gamma) [1-f(\gamma)]}{\f_\mathrm{j}(\gamma)-\f},
\label{eq:fv}
\end{equation}
where both $f(\gamma)$ and $\f_\mathrm{j}(\gamma)$ are extracted from fitting the simulation data for
$p\geq p_\mathrm{min}$ (see the table in~\cite{appendix}).
Very close to jamming ($p \gtrsim 10^5$), $f(\gamma)$ can be interpreted as the fraction of ``rattlers'' present,
but in the regime where the fluid first becomes non-ergodic, caging heterogeneity results in a larger effective $f$~\cite{donev:2005}. We find that with decreasing $\gamma$, $f(\gamma)$ converges to values of order $10\%$,
with only a weak $d$ dependence (Fig.~\ref{fig:comp}).

\paragraph*{Data analysis -}
The numerical results qualitatively agree with the RFOT scenario (see Ref.~\cite{parisi:2010} for details).
According to the theory, the glassy states for moderately small $\gamma$ should converge as a power law $\gamma^\alpha$ to a ``threshold'' glass that eventually jams at $\f_{\rm th}$.
The dynamical transition density $\f_{\rm d}$~\cite{footnoted}
separates the equilibrium fluid from this glass~\cite{MK11},
but at much slower compaction rates,
 $\gamma \lesssim\exp(-d)$, RFOT theory
also predicts that activated events allow the system to remain in
equilibrium up to higher densities. In this regime we expect a crossover to a logarithmic dependence of the glass EOS on $\gamma$, e.g., of the form $1/|\log(\gamma)|$.
As long as crystallization remains suppressed~\cite{skoge:2006,vanmeel:2009b},
an extrapolation to $\gamma\to 0$ in this regime allows to identify
a second order phase transition to an ``ideal'' glass, at a density
$\f_{\rm K}$ that corresponds to the Kauzmann point.
This ideal glass then jams at the glass close-packed (GCP) density $\f_{\mathrm{GCP}}$. Hence, from this mean-field
treatment, stable amorphous packings should exist in the interval $\f \in [\f_{\rm th},\f_{\rm GCP}]$~\cite{parisi:2010}.

\begin{figure}
\includegraphics[width=\columnwidth]{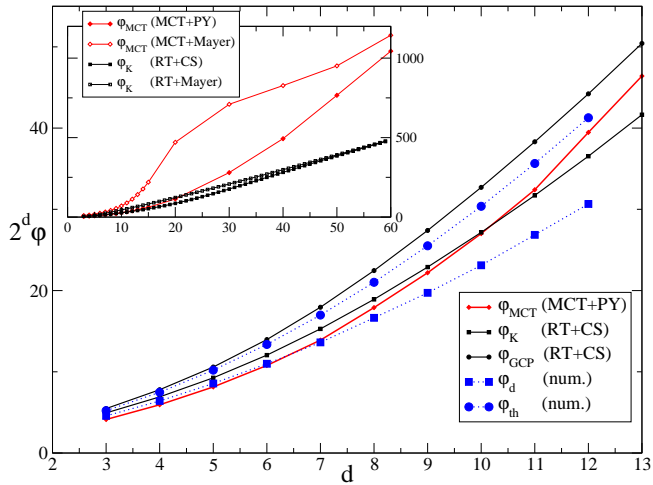}
\caption{
RFOT critical densities obtained from simulations, RT, and MCT. In the inset, the larger $d$ interval shows the approach to the asymptotic $d\to\infty$ limit, which
happens for $d\approx60$ in both RT and MCT.
}
\label{fig:phi}
\end{figure}

Based on this description, it is clear that the activated regime of RFOT is inaccessible in high dimensions. We therefore focus on the power-law regime and on the threshold glass.
In the inset of Fig.~\ref{fig:comp}, the reported $f(\gamma)$ and $\f_\mathrm{j}(\gamma)$ are both linear as
functions of
$\gamma^{\alpha}$ with $\alpha\approx0.5$.
The exponent is expected to weakly depend on dimensionality, especially at low $d$, but this value is within the
numerically reasonable range for all systems studied. Because the subsequent analysis is rather insensitive to the
precise value of $\alpha$, for simplicity it is kept constant. Extrapolating the results to $\gamma=0$ gives
the parameters $\f_{\rm th}$ and $f_{\rm th}$ for the threshold glass free volume EOS reported in Fig.~\ref{fig:comp}.
Its intersection with the fluid EOS then provides a numerical estimate for $\f_{\rm d}$.
These values are very close to previous numerical estimates in
$d=3$~\cite{brambilla:2009} and $d=4$~\cite{charbonneau:2010},
and the results for $\f_{\rm th}$ agree with the results of Refs.~\cite{skoge:2006,footnote:3}, which demonstrates the coherence of our analysis in low dimensions.
The numerical results reported in Table~\ref{tab:I} and plotted in Fig.~\ref{fig:phi} further
show that the dimensional evolution of $\f_{\mathrm{th}}$ and $\f_{\mathrm{d}}$ is smooth.
Interestingly, dimensions where the crystal structure is singularly dense,
such as $d=8$ and $d=12$~\cite{conway:1988}, do not present any echo of that singularity, 
which illustrates
the smooth $d$ dependence of the fluid structure.

\paragraph*{Theory -}
The numerical results can also be quantitatively compared with the analytical estimates for the RFOT critical densities. Although a reliable calculation of $\f_{\rm d}$ and $\f_{\rm th}$ has not yet been obtained within RT,
the latter gives results for $\f_{\rm K}$ and $\f_{\rm GCP}$~\cite{parisi:2010},
taking the Carnahan-Starling EOS with parameter $A_{d}^{\rm CS}$ as input.
As we mentioned above, the fitted values $A_{d}^{\rm fit}$ are slightly different, but that difference only weakly perturbs
the results. Because $A_{d}^{\rm fit}$ is only available for $d\leq 12$, we use instead $A_{d}^{\rm CS}$ from Ref.~\cite{parisi:2010}.
As expected from RFOT theory, $\f_{\rm K} > \f_{\rm d}$ and $\f_{\rm GCP} > \f_{\rm th}$, and the values
follow a similar trend with  $d$ (Table~\ref{tab:I}).

\begin{table}
\centering
\begin{tabular}{|c|cc|ccc|}
\hline
$d$ & $\f_{\rm d}$ & $\f_{\mathrm{th}}$ & $\f_\mathrm{K}$  & $\f_{\mathrm{GCP}}$ & $\f_{\mathrm{MCT}}$  \\
& &  & (RT+CS)     & (RT+CS)       & (MCT+PY)        \\
\hline
3        & 0.571     & 0.651             &  0.618       & 0.684          & 0.516      \\
4         & 0.401     & 0.467      &  0.432       & 0.487          & 0.371      \\
5        & 0.267       & 0.319          &  0.289       & 0.331          & 0.254      \\
6       & 0.172    & 0.209       &  0.189       & 0.219          & 0.169       \\
7       & 0.106          & 0.133                  &  0.120       & 0.140          & 0.109          \\
8     & 0.0648       & 0.0821                     &  0.0738       & 0.0875          & 0.0699          \\
9      & 0.0385        & 0.0498                     &  0.0447       & 0.0535          & 0.0434         \\
10     & 0.0226       & 0.0297             &  0.0266       & 0.0319          & 0.0264          \\
11       & 0.0131         & 0.0174                 &  0.0155       & 0.0187          & 0.0158         \\
12       & 0.0075      & 0.0101                &  0.00891       & 0.0108          & 0.00964         \\\hline
$\infty$       &              &             & $d \ln d \, 2^{-d}$ & $d \ln d \, 2^{-d}$ & $0.22 \, d^2 \, 2^{-d}$      \\
\hline
\end{tabular}
\caption{
Numerical values of the critical densities obtained from the simulations extrapolated at $\gamma=0$ and the RFOT theories. The uncertainty on the numerical results is on the last reported digit~\cite{appendix}.
}
\label{tab:I}
\end{table}

The MCT analysis follows the approach of Refs.~\cite{ikeda:2010,schmid:2010} using the Percus-Yevick (PY) structure factor calculated iteratively with a numerical Hankel transformation of order $d/2-1$.
Using instead the hypernetted chain (HNC) input for the structure
factor does not strongly affect the results. For the dynamical and static theories to be consistent,
the transition density $\f_{\rm MCT}$ predicted by MCT, reported in Table~\ref{tab:I}
and plotted in Fig.~\ref{fig:phi}, should coincide with $\f_{\rm d}$.
It is not the case. The MCT transition at $\f_{\rm MCT}$ increases too fast with $d$ and around $d=13$ will likely cross the numerical
estimate of $\f_{\rm th}$. This situation is paradoxical, because the liquid should fall out of equilibrium well {\it before} jamming occurs. The usual suggestion that activated events may improve the consistency
by increasing $\f_{\rm MCT}$ would here make things worse.

The inset in Fig.~\ref{fig:phi} presents $\f_{\rm MCT}$ and $\f_{\rm K}$ for larger dimensions. Both curves
approach the results obtained by neglecting all structure for the liquid~\cite{PS00}, which amounts
to using the van der Waals' expression
$p = 1 + 2^{d-1} \f$ for the fluid EOS and the Mayer function $\hat{f}(r) = -\theta(\sigma-r)$ for the direct correlation
function~\cite{ikeda:2010,schmid:2010}. This treatment is exact for $d\to\infty$, a fact that is also consistent with our simulation results for the fluid EOS. In this asymptotic large $d$ regime, MCT predicts $\f_{\rm MCT} \sim 0.22 d^2 \, 2^{-d}$~\cite{ikeda:2010,schmid:2010}, while
RT predicts $\f_{\rm K} \sim d\log(d) \, 2^{-d}$~\cite{parisi:2010}. The simulation results at intermediate $d$ thus give supporting evidence that the scaling
predicted by MCT is incorrect.

\paragraph*{Conclusions -}
Our numerical results extend previous estimates of the glass transition density~\cite{brambilla:2009,charbonneau:2010}
and of the amorphous packing density~\cite{skoge:2006} up to $d=12$. These results are obtained partly thanks to methodological improvements~\cite{donev:2005b} over the approach of Ref.~\cite{skoge:2006}. Reassuringly, our procedure for the data analysis, which is grounded in the RFOT scenario~\cite{parisi:2010},
produces results that are consistent with estimates obtained through alternate routes. Two key features arise from the work.
First, our numerical estimates of $\f_{\rm th}$ are slightly smaller than the values of $\f_{\rm GCP}$ predicted by RT and follow a similar trend,
indicating that RT provides a reliable prediction for the jamming density of hard spheres in high dimensions. In particular, the random close-packed
state of frictionless hard spheres, which is expected to be protocol dependent~\cite{KK07,parisi:2010}, should fall within the range of densities predicted by the
RFOT scenario $\f \in [\f_{\rm th},\f_{\rm GCP}]$. We thus expect our numerical results for $\f_{\rm th}$ to be good estimates of that density,
irrespective of the protocol used. Second, our results for the dynamical (glass) transition $\f_{\rm d}$
are slightly smaller than the ideal glass (Kauzmann) transition $\f_{\rm K}$ predicted by RT, as it should be. But at the same time, the MCT
prediction $\f_{\rm MCT}$ does not coincide with $\f_{\rm d}$, unlike what the RFOT scenario suggests. The two quantities additionally show very different
trends with dimension and the discrepancy increases at large $d$, which suggests that the standard MCT formulation~\cite{gotze:2009,ikeda:2010,schmid:2010} is not exact at large $d$. Echoing and amplifying the call of Ref.~\cite{ikeda:2010}, a new theory should be constructed, in order to interpolate between
MCT in $d=3$ and the large $d$ mean-field regime.

Whether this theory will
be a minor modification of MCT or a completely new theory remains an
open question.
Indeed, mean-field models displaying a RFOT are described by MCT-like
equations~\cite{kirkpatrick:1989,kirkpatrick:1987}, and MCT itself has been described as a type of Landau theory of the dynamical
transition~\cite{bouchaud:2010,andreanov:2009}. It is thus possible that a modified MCT with a similar structure might be exact in large $d$.
Steps in this direction have been already taken~\cite{MK11,jacquin:2011}. Constructing a complete RFOT theory that includes both RT and a
modified MCT would be an important intellectual advance, as it would be the first complete theory of the glass and jamming transitions for particle-based systems.
This advance would open the way towards a more systematic construction of RFOT, through a large $d$ expansion or a renormalization group approach. Some of the approximations underlying the construction of MCT might also be better justified, leading to a fully predictive glass theory for $d\leq3$.

\acknowledgments
We thank J.~Kurchan, R.~Mari, K.~Miyazaki and R.~Schilling for
stimulating
discussions.
PC acknowledges NSF support No. DMR-1055586.
%

\begin{widetext}

\section{Appendix}

\begin{table*}[h]
\centering
\begin{tabular}{|c|cccc|cll|ccc|}
\hline
$d$ &  $N$        & $A_{d}^{\rm fit}$ &  $A_{d}^{\rm CS}$ &  $1/p_{\rm min}$ & $f_{\rm th}$    & $2^d \f_{\rm d}$ & $2^d \f_{\mathrm{th}}$        & $2^d \f_\mathrm{K}$  & $2^d \f_{\mathrm{GCP}}$ & $2^d \f_{\mathrm{MCT}}$  \\
     &            &                       &  Ref.~\cite{SMS89}  &                 & & &  & (RT+CS)     & (RT+CS)       & (MCT+PY)        \\
\hline
3         & 8000       & 0.5                   &  0.5                  &  0.03          & 0.11(2)   & 4.57(6)      & 5.21(1)              &  4.94       & 5.47          & 4.13      \\
4         & 8000       & 0.114                 &  -0.051               &  0.02      & 0.12(1)        & 6.41(6)      & 7.48(1)       &  6.91       & 7.79          & 5.94      \\
5        & 8000       & -1.086                &  -1.625               &  0.02       & 0.124(5)     & 8.56(5)        & 10.20(1)           &  9.26       & 10.6          & 8.13      \\
6       & 8000       & -3.900                &  -4.910               &  0.02        & 0.128(6)       & 10.99(8)     & 13.37(2)        &  12.1       & 14.0          & 10.8       \\
7        & 8000       & -9.294                &  -11.06               &  0.02        & 0.130(4)      & 13.6(1)          & 16.98(4)                  &  15.3       & 17.9          & 13.9          \\
\hline
8     & 1400       & -19.38                &  -22.03               &  0.02            & 0.14(2)    & 16.5(3)       & 20.92(5)                     &  18.9       & 22.4          & 17.9          \\
8     & 2500       & -19.38                &                       &  0.02            & 0.136(7)    & 16.6(2)       & 21.02(5)                     &             &               &               \\
8     & 8000       & -19.38                &                       &  0.02            & 0.135(4)    & 16.6(1)       & 21.00(4)                     &             &               &               \\
8     & 32768       & -19.38                &                       &  0.02            & 0.133(3)    & 16.6(1)       & 21.02(5)                     &             &               &               \\
\hline
9       & 8000       & -35.46                &  -41.10               &  0.02          & 0.137(4)     & 19.7(2)        & 25.51(7)                     &  22.9       & 27.4          & 22.2         \\
10      & 20000      & -62.50                &  -73.83               &  0.02            & 0.139(4)        & 23.1(2)       & 30.37(8)              &  27.2       & 32.7          & 27.0          \\
11       & 50000      & -107.0                &  -129.6               &  0.015         & 0.141(1)       & 26.9(2)          & 35.64(9)                  &  31.7       & 38.3          & 32.4         \\
12        & 125000     & -176.0                &  -224.3               &  0.015           & 0.145(6)      & 30.6(5)         & 41.3(2)                &  36.5       & 44.2          & 39.5         \\
\hline
$\infty$ &            &                       &                       &                 &              &              &             & $d \ln d$ & $d \ln d$ & $0.22 \, d^2$      \\
\hline
\end{tabular}
\caption{
Numerical values of the RFOT critical densities and of the simulation parameters. The simulation results are extrapolated at $\gamma=0$ according to the procedure discussed in the text.
}
\label{tab:II}
\end{table*}

\paragraph*{Error analysis -}
From the simulation results, the error on $A_{d}^{\rm fit}$ is found to be very small and, for all practical purposes, the quantity is here considered exact. While the error on $\f_{\rm th}$ and $f_{\mathrm{th}}$ are estimated by taking three times the error on the linear regression coefficients, for $\f_{\rm d}$, a standard error propagation is conducted
starting from its definition. Indeed $\f_{\rm d}$ is the intersection of the fluid EOS and the threshold glass EOS defined by the extrapolation
of the free volume equation to $\gamma=0$:
\beq
p_{\rm fluid}(\f) = p_{\rm fv}(\gamma\to 0,\f) \equiv \frac{d \f_{\mathrm{th}} (1-f_{\mathrm{th}})}{\f_{\mathrm{th}} - \f}  \ .
\eeq
Defining $q(\f)=1/p_{\rm fluid}(\f)$, we then get that
\begin{equation}
\delta \f_{\rm d}=\left|\frac{1-d \, (1-f_{\mathrm{th}}) \, q(\f_{\rm d})]}{1+d \, (1-f_{\mathrm{th}}) \, \f_{\rm th} \, q'(\f_{\rm d})}\right|\delta \f_{\mathrm{th}}+\left|\frac{d \, \f_{\rm th} \, q(\f_{\rm d})}
{1+d \, (1-f_{\mathrm{th}}) \, \f_{\rm th} \, q'(\f_{\rm d})}\right|\delta f_{\mathrm{th}} .
\end{equation}

\paragraph*{Finite-size effects - }
The finite-size arguments that we presented in the main text are
tested by comparing $d=8$ systems with $N=1.4$K, $2.5$K, $8$K and $32$K.
Although the number of particles increases by nearly $50\%$ per dimension over this sequence, the compaction curves are almost indistinguishable except for a slight finite size effect for the largest $\gamma=0.1$.
The results of our data analysis coincide within numerical errors, estimated as discussed in the previous
paragraph (Table~\ref{tab:II}).

\end{widetext}


\vskip-10pt

\begin{thebibliography}{33}%
\makeatletter
\providecommand \@ifxundefined [1]{%
 \@ifx{#1\undefined}
}%
\providecommand \@ifnum [1]{%
 \ifnum #1\expandafter \@firstoftwo
 \else \expandafter \@secondoftwo
 \fi
}%
\providecommand \@ifx [1]{%
 \ifx #1\expandafter \@firstoftwo
 \else \expandafter \@secondoftwo
 \fi
}%
\providecommand \natexlab [1]{#1}%
\providecommand \enquote  [1]{``#1''}%
\providecommand \bibnamefont  [1]{#1}%
\providecommand \bibfnamefont [1]{#1}%
\providecommand \citenamefont [1]{#1}%
\providecommand \href@noop [0]{\@secondoftwo}%
\providecommand \href [0]{\begingroup \@sanitize@url \@href}%
\providecommand \@href[1]{\@@startlink{#1}\@@href}%
\providecommand \@@href[1]{\endgroup#1\@@endlink}%
\providecommand \@sanitize@url [0]{\catcode `\\12\catcode `\$12\catcode
  `\&12\catcode `\#12\catcode `\^12\catcode `\_12\catcode `\%12\relax}%
\providecommand \@@startlink[1]{}%
\providecommand \@@endlink[0]{}%
\providecommand \url  [0]{\begingroup\@sanitize@url \@url }%
\providecommand \@url [1]{\endgroup\@href {#1}{\urlprefix }}%
\providecommand \urlprefix  [0]{URL }%
\providecommand \Eprint [0]{\href }%
\providecommand \doibase [0]{http://dx.doi.org/}%
\providecommand \selectlanguage [0]{\@gobble}%
\providecommand \bibinfo  [0]{\@secondoftwo}%
\providecommand \bibfield  [0]{\@secondoftwo}%
\providecommand \translation [1]{[#1]}%
\providecommand \BibitemOpen [0]{}%
\providecommand \bibitemStop [0]{}%
\providecommand \bibitemNoStop [0]{.\EOS\space}%
\providecommand \EOS [0]{\spacefactor3000\relax}%
\providecommand \BibitemShut  [1]{\csname bibitem#1\endcsname}%
\let\auto@bib@innerbib\@empty
\bibitem [{\citenamefont {Anderson}(1995)}]{anderson:1995}%
  \BibitemOpen
  \bibfield  {author} {\bibinfo {author} {\bibfnamefont {P.~W.}\ \bibnamefont
  {Anderson}},\ }\href@noop {} {\bibfield  {journal} {\bibinfo  {journal}
  {Science}\ }\textbf {\bibinfo {volume} {267}},\ \bibinfo {pages} {1615}
  (\bibinfo {year} {1995})}\BibitemShut {NoStop}%
\bibitem [{\citenamefont {Charbonneau}\ \emph {et~al.}(2010)\citenamefont
  {Charbonneau}, \citenamefont {Ikeda}, \citenamefont {van Meel},\ and\
  \citenamefont {Miyazaki}}]{charbonneau:2010}%
  \BibitemOpen
  \bibfield  {author} {\bibinfo {author} {\bibfnamefont {P.}~\bibnamefont
  {Charbonneau}}, \bibinfo {author} {\bibfnamefont {A.}~\bibnamefont {Ikeda}},
  \bibinfo {author} {\bibfnamefont {J.~A.}\ \bibnamefont {van Meel}}, \ and\
  \bibinfo {author} {\bibfnamefont {K.}~\bibnamefont {Miyazaki}},\ }\href@noop
  {} {\bibfield  {journal} {\bibinfo  {journal} {Phys. Rev. E}\ }\textbf
  {\bibinfo {volume} {81}},\ \bibinfo {pages} {040501(R)} (\bibinfo {year}
  {2010})}\BibitemShut {NoStop}%
\bibitem [{\citenamefont {Conway}\ and\ \citenamefont
  {Sloane}(1988)}]{conway:1988}%
  \BibitemOpen
  \bibfield  {author} {\bibinfo {author} {\bibfnamefont {J.~H.}\ \bibnamefont
  {Conway}}\ and\ \bibinfo {author} {\bibfnamefont {N.~J.~A.}\ \bibnamefont
  {Sloane}},\ }\href@noop {} {\emph {\bibinfo {title} {Sphere Packings,
  Lattices and Groups}}}\ (\bibinfo  {publisher} {Springer-Verlag},\ \bibinfo
  {address} {New York},\ \bibinfo {year} {1988});
  H.~Cohn, Proceedings of the international congress of mathematicians (ICM 2010), 2416 ({\tt arXiv:1003.3053})\BibitemShut {NoStop}%
\bibitem [{\citenamefont {Kirkpatrick}\ \emph {et~al.}(1989)\citenamefont
  {Kirkpatrick}, \citenamefont {Thirumalai},\ and\ \citenamefont
  {Wolynes}}]{kirkpatrick:1989}%
  \BibitemOpen
  \bibfield  {author} {\bibinfo {author} {\bibfnamefont {T.~R.}\ \bibnamefont
  {Kirkpatrick}}, \bibinfo {author} {\bibfnamefont {D.}~\bibnamefont
  {Thirumalai}}, \ and\ \bibinfo {author} {\bibfnamefont {P.~G.}\ \bibnamefont
  {Wolynes}},\ }\href@noop {} {\bibfield  {journal} {\bibinfo  {journal} {Phys.
  Rev. A}\ }\textbf {\bibinfo {volume} {40}},\ \bibinfo {pages} {1045}
  (\bibinfo {year} {1989})}\BibitemShut {NoStop}%
\bibitem [{\citenamefont {G\"otze}(2009)}]{gotze:2009}%
  \BibitemOpen
  \bibfield  {author} {\bibinfo {author} {\bibfnamefont {W.}~\bibnamefont
  {G\"otze}},\ }\href@noop {} {\emph {\bibinfo {title} {Complex Dynamics of
  Glass-Forming Liquids}}},\ \bibinfo {series} {International Series of
  Monographs on Physics}, Vol.\ \bibinfo {volume} {143}\ (\bibinfo  {publisher}
  {Oxford University Press},\ \bibinfo {address} {Oxford},\ \bibinfo {year}
  {2009})\BibitemShut {NoStop}%
\bibitem [{\citenamefont {M\'ezard}\ and\ \citenamefont
  {Parisi}(1999)}]{mezard:1999}%
  \BibitemOpen
  \bibfield  {author} {\bibinfo {author} {\bibfnamefont {M.}~\bibnamefont
  {M\'ezard}}\ and\ \bibinfo {author} {\bibfnamefont {G.}~\bibnamefont
  {Parisi}},\ }\href@noop {} {\bibfield  {journal} {\bibinfo  {journal} {Phys.
  Rev. Lett.}\ }\textbf {\bibinfo {volume} {82}},\ \bibinfo {pages} {747}
  (\bibinfo {year} {1999})}\BibitemShut {NoStop}%
\bibitem [{\citenamefont {Kirkpatrick}\ and\ \citenamefont
  {Wolynes}(1987)}]{kirkpatrick:1987}%
  \BibitemOpen
  \bibfield  {author} {\bibinfo {author} {\bibfnamefont {T.~R.}\ \bibnamefont
  {Kirkpatrick}}\ and\ \bibinfo {author} {\bibfnamefont {P.~G.}\ \bibnamefont
  {Wolynes}},\ }\href@noop {} {\bibfield  {journal} {\bibinfo  {journal} {Phys.
  Rev. A}\ }\textbf {\bibinfo {volume} {35}},\ \bibinfo {pages} {3072}
  (\bibinfo {year} {1987})}\BibitemShut {NoStop}%
\bibitem [{\citenamefont {Szamel}(2010)}]{szamel:2010}%
  \BibitemOpen
  \bibfield  {author} {\bibinfo {author} {\bibfnamefont {G.}~\bibnamefont
  {Szamel}},\ }\href@noop {} {\bibfield  {journal} {\bibinfo  {journal}
  {Europhys. Lett.}\ }\textbf {\bibinfo {volume} {91}},\ \bibinfo {pages}
  {56004} (\bibinfo {year} {2010})}\BibitemShut {NoStop}%
\bibitem [{\citenamefont {Schmid}\ and\ \citenamefont
  {Schilling}(2010)}]{schmid:2010}%
  \BibitemOpen
  \bibfield  {author} {\bibinfo {author} {\bibfnamefont {B.}~\bibnamefont
  {Schmid}}\ and\ \bibinfo {author} {\bibfnamefont {R.}~\bibnamefont
  {Schilling}},\ }\href@noop {} {\bibfield  {journal} {\bibinfo  {journal}
  {Phys. Rev. E}\ }\textbf {\bibinfo {volume} {81}},\ \bibinfo {pages} {041502}
  (\bibinfo {year} {2010})};\ \bibfield  {author} {\bibinfo {author} {\bibfnamefont {R.}~\bibnamefont
  {Schilling}}\ and\ \bibinfo {author} {\bibfnamefont {B.}~\bibnamefont
  {Schmid}},\ }\href@noop {} {\bibfield  {journal} {\bibinfo  {journal} {Phys.
  Rev. Lett.}\ }\textbf {\bibinfo {volume} {106}},\ \bibinfo {pages} {049601}
  (\bibinfo {year} {2011})}\BibitemShut {NoStop}%
\bibitem [{\citenamefont {Ikeda}\ and\ \citenamefont
  {Miyazaki}(2010)}]{ikeda:2010}%
  \BibitemOpen
  \bibfield  {author} {\bibinfo {author} {\bibfnamefont {A.}~\bibnamefont
  {Ikeda}}\ and\ \bibinfo {author} {\bibfnamefont {K.}~\bibnamefont
  {Miyazaki}},\ }\href@noop {} {\bibfield  {journal} {\bibinfo  {journal}
  {Phys. Rev. Lett.}\ }\textbf {\bibinfo {volume} {104}},\ \bibinfo {pages}
  {255704} (\bibinfo {year} {2010})};\ {\bibfield  {journal} {\bibinfo  {journal}
  {Phys. Rev. Lett.}\ }\textbf {\bibinfo {volume} {106}},\ \bibinfo {pages}
  {049602} (\bibinfo {year} {2011})}\BibitemShut {NoStop}%
\bibitem [{\citenamefont {Parisi}\ and\ \citenamefont
  {Zamponi}(2010)}]{parisi:2010}%
  \BibitemOpen
  \bibfield  {author} {\bibinfo {author} {\bibfnamefont {G.}~\bibnamefont
  {Parisi}}\ and\ \bibinfo {author} {\bibfnamefont {F.}~\bibnamefont
  {Zamponi}},\ }\href@noop {} {\bibfield  {journal} {\bibinfo  {journal} {Rev.
  Mod. Phys.}\ }\textbf {\bibinfo {volume} {82}},\ \bibinfo {pages} {789}
  (\bibinfo {year} {2010})}\BibitemShut {NoStop}%
\bibitem [{\citenamefont {Bouchaud}(2010)}]{bouchaud:2010}%
  \BibitemOpen
  \bibfield  {author} {\bibinfo {author} {\bibfnamefont {J.-P.}\ \bibnamefont
  {Bouchaud}},\ }\href@noop {} {\enquote
  {\bibinfo {title} {The mode-coupling theory of supercooled liquids: Does it
  wear any clothes?}}\ } {http://www.condmatjournalclub.org/?p=1022}  (\bibinfo {year} {2010})\BibitemShut {NoStop}%
\bibitem{KS91}
W.~Kob and R.~Schilling, J.~Phys.:~Condens.~Matter {\bf 3}, 9195 (1991).
\bibitem [{\citenamefont {Torquato}\ and\ \citenamefont
  {Stillinger}(2006)}]{torquato:2006b}%
  \BibitemOpen
  \bibfield  {author} {\bibinfo {author} {\bibfnamefont {S.}~\bibnamefont
  {Torquato}}\ and\ \bibinfo {author} {\bibfnamefont {F.~H.}\ \bibnamefont
  {Stillinger}},\ }\href@noop {} {\bibfield  {journal} {\bibinfo  {journal}
  {Exp. Math.}\ }\textbf {\bibinfo {volume} {15}},\ \bibinfo {pages} {307}
  (\bibinfo {year} {2006})}\BibitemShut {NoStop}%
\bibitem [{\citenamefont {van Meel}\ \emph {et~al.}(2009)\citenamefont {van
  Meel}, \citenamefont {Charbonneau}, \citenamefont {Fortini},\ and\
  \citenamefont {Charbonneau}}]{vanmeel:2009b}%
  \BibitemOpen
  \bibfield  {author} {\bibinfo {author} {\bibfnamefont {J.~A.}\ \bibnamefont
  {van Meel}}, \bibinfo {author} {\bibfnamefont {B.}~\bibnamefont
  {Charbonneau}}, \bibinfo {author} {\bibfnamefont {A.}~\bibnamefont
  {Fortini}}, \ and\ \bibinfo {author} {\bibfnamefont {P.}~\bibnamefont
  {Charbonneau}},\ }\href@noop {} {\bibfield  {journal} {\bibinfo  {journal}
  {Phys. Rev. E}\ }\textbf {\bibinfo {volume} {80}},\ \bibinfo {pages} {061110}
  (\bibinfo {year} {2009})}\BibitemShut {NoStop}%
\bibitem [{\citenamefont {Song}\ \emph {et~al.}(2008)\citenamefont {Song},
  \citenamefont {Wang},\ and\ \citenamefont {Makse}}]{song:2008}%
  \BibitemOpen
  \bibfield  {author} {\bibinfo {author} {\bibfnamefont {C.}~\bibnamefont
  {Song}}, \bibinfo {author} {\bibfnamefont {P.}~\bibnamefont {Wang}}, \ and\
  \bibinfo {author} {\bibfnamefont {H.~A.}\ \bibnamefont {Makse}},\ }\href@noop
  {} {\bibfield  {journal} {\bibinfo  {journal} {Nature}\ }\textbf {\bibinfo
  {volume} {453}},\ \bibinfo {pages} {629} (\bibinfo {year}
  {2008})};\ \bibfield  {author} {\bibinfo {author} {\bibfnamefont {Y.}~\bibnamefont
  {Jin}\ \emph{et~al.}\ }} 
  \href@noop {} {\bibfield
  {journal} {\bibinfo  {journal} {Phys. Rev. E}\ }\textbf {\bibinfo {volume}
  {82}},\ \bibinfo {pages} {051126} (\bibinfo {year} {2010})}\BibitemShut {NoStop}%
\bibitem [{\citenamefont {Skoge}\ \emph {et~al.}(2006)\citenamefont {Skoge},
  \citenamefont {Donev}, \citenamefont {Stillinger},\ and\ \citenamefont
  {Torquato}}]{skoge:2006}%
  \BibitemOpen
  \bibfield  {author} {\bibinfo {author} {\bibfnamefont {M.}~\bibnamefont
  {Skoge}}, \bibinfo {author} {\bibfnamefont {A.}~\bibnamefont {Donev}},
  \bibinfo {author} {\bibfnamefont {F.~H.}\ \bibnamefont {Stillinger}}, \ and\
  \bibinfo {author} {\bibfnamefont {S.}~\bibnamefont {Torquato}},\ }\href@noop
  {} {\bibfield  {journal} {\bibinfo  {journal} {Phys. Rev. E}\ }\textbf
  {\bibinfo {volume} {74}},\ \bibinfo {pages} {041127} (\bibinfo {year}
  {2006})}\BibitemShut {NoStop}%
\bibitem [{foo({\natexlab{a}})}]{footnote:1}%
  \BibitemOpen {In contrast to the
  original implementation, a canonical definition of temperature is here used,
  which rescales the time unit $\sqrt{\beta m\sigma^2}$ for particles of unit mass
  $m$.}\BibitemShut {Stop}%
\bibitem{donev:2005b}
L.~Verlet, Phys. Rev. {\bf 159}, 98 (1967);
A.~Donev, S.~Torquato, F.~H.~Stillinger, J.~Comp.~Phys. {\bf 202}, 765 (2005).
\bibitem [{\citenamefont {Parisi}\ and\ \citenamefont {Slanina}(2000)}]{PS00}%
  \BibitemOpen
  \bibfield  {author} {\bibinfo {author} {\bibfnamefont {G.}~\bibnamefont
  {Parisi}}\ and\ \bibinfo {author} {\bibfnamefont {F.}~\bibnamefont
  {Slanina}},\ }\href@noop {} {\bibfield  {journal}
  {\bibinfo  {journal} {Phys. Rev. E}\ }\textbf {\bibinfo {volume} {62}},\
  \bibinfo {pages} {6554} (\bibinfo {year} {2000})}\BibitemShut {NoStop}%
  \bibitem{appendix}
See Appendix to this paper for numerical details, finite-size studies, and error analysis. 
\bibitem [{\citenamefont {Parisi}(1998)}]{Pa98}%
  \BibitemOpen
  \bibfield  {author} {\bibinfo {author} {\bibfnamefont {G.}~\bibnamefont
  {Parisi}},\ }\href@noop {} {\emph {\bibinfo {title} {{Statistical Field
  Theory}}}}\ (\bibinfo  {publisher} {Perseus Group},\ \bibinfo {year}
  {1998})\BibitemShut {NoStop}%
\bibitem [{foo({\natexlab{b}})}]{footnote:2}%
  \BibitemOpen {The many virial
  coefficients known for hard spheres, e.g., Ref.~\cite{bishop:2008},
  insufficiently capture the equation of state near the onset density of the
  non-ergodic regime.}\BibitemShut {Stop}%
\bibitem [{\citenamefont {Song}\ \emph {et~al.}(1989)\citenamefont {Song},
  \citenamefont {Mason},\ and\ \citenamefont {Stratt}}]{SMS89}%
  \BibitemOpen
  \bibfield  {author} {\bibinfo {author} {\bibfnamefont {Y.}~\bibnamefont
  {Song}}, \bibinfo {author} {\bibfnamefont {E.~A.}\ \bibnamefont {Mason}}, \
  and\ \bibinfo {author} {\bibfnamefont {R.~M.}\ \bibnamefont {Stratt}},\
  }\href@noop {} {\bibfield  {journal} {\bibinfo  {journal} {J. Phys. Chem.}\
  }\textbf {\bibinfo {volume} {93}},\ \bibinfo {pages} {6916} (\bibinfo {year}
  {1989})}\BibitemShut {NoStop}%
\bibitem [{\citenamefont {Donev}\ \emph {et~al.}(2005)\citenamefont {Donev},
  \citenamefont {Torquato},\ and\ \citenamefont {Stillinger}}]{donev:2005}%
  \BibitemOpen
  \bibfield  {author} {\bibinfo {author} {\bibfnamefont {A.}~\bibnamefont
  {Donev}}, \bibinfo {author} {\bibfnamefont {S.}~\bibnamefont {Torquato}}, \
  and\ \bibinfo {author} {\bibfnamefont {F.~H.}\ \bibnamefont {Stillinger}},\
  }\href@noop {} {\bibfield  {journal} {\bibinfo  {journal} {Phys. Rev. E}\
  }\textbf {\bibinfo {volume} {71}},\ \bibinfo {pages} {011105} (\bibinfo
  {year} {2005})}\BibitemShut {NoStop}%
\bibitem [{\citenamefont {Kamien}\ and\ \citenamefont
  {Liu}(2007)}]{kamien:2007}%
  \BibitemOpen
  \bibfield  {author} {\bibinfo {author} {\bibfnamefont {R.~D.}\ \bibnamefont
  {Kamien}}\ and\ \bibinfo {author} {\bibfnamefont {A.~J.}\ \bibnamefont
  {Liu}},\ }\href@noop {} {\bibfield  {journal} {\bibinfo  {journal} {Phys.
  Rev. Lett.}\ }\textbf {\bibinfo {volume} {99}},\ \bibinfo {pages} {155501}
  (\bibinfo {year} {2007})}\BibitemShut {NoStop}%
\bibitem{footnoted}
We use the canonical notation for the dynamical transition $\varphi_\mathrm{d}$, but the reader should not confuse this ‘‘d’’ subscript with dimensionality $d$.
\bibitem [{\citenamefont {Mari}\ and\ \citenamefont {Kurchan}(2011)}]{MK11}%
  \BibitemOpen
  \bibfield  {author} {\bibinfo {author} {\bibfnamefont {R.}~\bibnamefont
  {Mari}}\ and\ \bibinfo {author} {\bibfnamefont {J.}~\bibnamefont {Kurchan}},\
  }\href@noop {} {\bibfield  {journal} {\bibinfo  {journal} {{\tt
  arXiv:1104.3420}}\ } (\bibinfo {year} {2011})}\BibitemShut {NoStop}%
\bibitem [{\citenamefont {Brambilla}\ \emph {et~al.}(2009)\citenamefont
  {Brambilla}, \citenamefont {El~Masri}, \citenamefont {Pierno}, \citenamefont
  {Berthier}, \citenamefont {Cipelletti}, \citenamefont {Petekidis},\ and\
  \citenamefont {Schofield}}]{brambilla:2009}%
  \BibitemOpen
  \bibfield  {author} {\bibinfo {author} {\bibfnamefont {G.}~\bibnamefont
  {Brambilla}}\ \emph{et~al.}\ }, 
Phys. Rev. Lett. {\bf 102}, {085703} ({2009})%
\bibitem [{foo({\natexlab{c}})}]{footnote:3}%
  \BibitemOpen
  {Within RFOT, the
  ``maximally random jammed'' packing discussed in Ref.~\cite{skoge:2006}
  should be very close to $\varphi_{\rm th}$~\cite{parisi:2010}.}\BibitemShut
  {Stop}%
\bibitem [{\citenamefont {Krzakala}\ and\ \citenamefont
  {Kurchan}(2007)}]{KK07}%
  \BibitemOpen
  \bibfield  {author} {\bibinfo {author} {\bibfnamefont {F.}~\bibnamefont
  {Krzakala}}\ and\ \bibinfo {author} {\bibfnamefont {J.}~\bibnamefont
  {Kurchan}},\ }\href@noop {} {\bibfield  {journal} {\bibinfo  {journal} {Phys.
  Rev. E}\ }\textbf {\bibinfo {volume} {76}},\ \bibinfo {pages} {021122}
  (\bibinfo {year} {2007})}; %
  S. Torquato and F. H. Stillinger, Rev. Mod. Phys. {\bf 82}, 2633 (2010).
\bibitem [{\citenamefont {Andreanov}\ \emph {et~al.}(2009)\citenamefont
  {Andreanov}, \citenamefont {Biroli},\ and\ \citenamefont
  {Bouchaud}}]{andreanov:2009}%
  \BibitemOpen
  \bibfield  {author} {\bibinfo {author} {\bibfnamefont {A.}~\bibnamefont
  {Andreanov}}, \bibinfo {author} {\bibfnamefont {G.}~\bibnamefont {Biroli}}, \
  and\ \bibinfo {author} {\bibfnamefont {J.-P.}\ \bibnamefont {Bouchaud}},\
  }\href@noop {} {\bibfield  {journal} {\bibinfo  {journal} {Europhys. Lett.}\
  }\textbf {\bibinfo {volume} {88}},\ \bibinfo {pages} {16001} (\bibinfo {year}
  {2009})}\BibitemShut {NoStop}%
\bibitem [{\citenamefont {Jacquin}\ and\ \citenamefont {van
  Wijland}(2011)}]{jacquin:2011}%
  \BibitemOpen
  \bibfield  {author} {\bibinfo {author} {\bibfnamefont {H.}~\bibnamefont
  {Jacquin}}\ and\ \bibinfo {author} {\bibfnamefont {F.}~\bibnamefont {van
  Wijland}},\ }\href@noop {}  {\bibfield
  {journal} {\bibinfo  {journal} {Phys. Rev. Lett.}\ }\textbf {\bibinfo
  {volume} {106}},\ \bibinfo {pages} {210602} (\bibinfo {year}
  {2011})}; Y.~Elskens and H.~L.~Frisch, Phys. Rev. A {\bf 37}, 4351 (1988) \BibitemShut {NoStop}%
\bibitem [{\citenamefont {Bishop}\ \emph {et~al.}(2008)\citenamefont {Bishop},
  \citenamefont {Clisby},\ and\ \citenamefont {Whitlock}}]{bishop:2008}%
  \BibitemOpen
  \bibfield  {author} {\bibinfo {author} {\bibfnamefont {M.}~\bibnamefont
  {Bishop}}, \bibinfo {author} {\bibfnamefont {N.}~\bibnamefont {Clisby}}, \
  and\ \bibinfo {author} {\bibfnamefont {P.~A.}\ \bibnamefont {Whitlock}},\
  }\href@noop {} {\bibfield  {journal} {\bibinfo  {journal} {J. Chem. Phys.}\
  }\textbf {\bibinfo {volume} {128}},\ \bibinfo {pages} {034506} (\bibinfo
  {year} {2008})}\BibitemShut {NoStop}%
\end{thebibliography}
\end{document}